\documentstyle[12pt]{article}

\begin{document}

\newcommand{\eq}[1]{eq.~(\ref{#1})}
\newcommand{\eqs}[2]{eqs.(\ref{#1}, \ref{#2})}
\newcommand{\eqss}[3]{eqs.(\ref{#1}, \ref{#2}, ref{#3})}
\newcommand{\ur}[1]{~(\ref{#1})}
\newcommand{\urs}[2]{~(\ref{#1},~\ref{#2})}
\newcommand{\urss}[3]{~(\ref{#1},~\ref{#2},~\ref{#3})}
\newcommand{\Eq}[1]{Eq.~(\ref{#1})}
\newcommand{\Eqs}[2]{Eqs.(\ref{#1},\ref{#2})}
\newcommand{\fig}[1]{Fig.~\ref{#1}}
\newcommand{\figs}[2]{Figs.\ref{#1},\ref{#2}}
\newcommand{\figss}[3]{Figs.\ref{#1},\ref{#2},\ref{#3}}
\newcommand{\beq}{\begin{equation}}
\newcommand{\eeq}{\end{equation}}
\newcommand{\e}{\varepsilon}
\newcommand{\ee}{\epsilon}
\newcommand{\bI}{\bar I}
\newcommand{\eoe}{(\bar{\eta}O\eta)}
\newcommand{\ui}{U_{int}(R,\rho_{1,2},O_{1,2})}
\newcommand{\thb}[3]{\bar{\eta}^{#1}_{#2 #3}}
\newcommand{\tha}[3]{\eta^{#1}_{#2 #3}}
\newcommand{\la}[1]{\label{#1}}
\newcommand{\YM}{Yang--Mills~}
\newcommand{\N}{$N_{CS}\;$}
\newcommand{\Nz}{$N_{CS}=0\;$}
\newcommand{\No}{$N_{CS}=1\;$}
\newcommand{\x}{{\bf x}}
\newcommand{\r}{{\bf r}}
\newcommand{\IIs}{$I$'s and ${\bar I}$'s }
\newcommand{\II}{$I{\bar I}$ }

\vspace{1.5cm}
\begin{center} \Large\bf
Clustering of Monopoles in the Instanton Vacuum
\end{center}
\vspace{.7cm}

\begin{center} {\large 
M.~Fukushima$^a$, S.~Sasaki$^a$, H.~Suganuma$^a$, 
A.~Tanaka$^a$, H.~Toki$^a$ \\
and D.~Diakonov$^b$} \\
\vskip .5true cm
$^a${\small\it Research Center for Nuclear Physics (RCNP), Osaka
University, Ibaraki, Osaka 567, Japan}
\vskip .5true cm
$^b${\small\it
Petersburg Nuclear Physics Institute, Gatchina, St.Petersburg 188350,
Russia}
\end{center}

\vskip 1true cm

\abstract{We generate a random instanton vacuum with 
various densities and size distributions. We perform numerically the 
maximally abelian gauge fixing of these configurations in order to find 
monopole trajectories induced by instantons. We find that 
instanton-induced monopole loops form enormous clusters 
occupying the whole physical volume, 
provided instantons are sufficiently dense. It indicates that 
confinement might be caused by instantons.}

\vskip 1.5true cm

\noindent
{\bf 1. Introduction} 

It is widely discussed that confinement, characterized by the area 
behavior of the Wilson loop, is related to 
monopoles~\cite{Nambu}-\cite{Ezawa}. 
There is also growing evidence that instantons play a key role 
in the nonperturbative QCD. Instantons should be present in the
Yang-Mills vacuum if only because a non-zero topological
susceptibility is needed to resolve the so-called $U_A(1)$
paradox~\cite{tH,Akas}. Also, instantons are playing an
important role in all phenomena related to chiral symmetry
breaking~\cite{D1}. Until now, however, there has been no evidence
that instantons have anything to do with confinement. Rather on the
contrary: the string tension is known to decrease with lattice
cooling, whereas instantons survive many cooling
steps~\cite{PV}-\cite{CGHN}.

It may seem thus that instantons and monopoles are representing
completely different sectors of physics, and are hardly related to one
another. Recently, however, there has been evidence indicating that
there might be a relation between the two
objects~\cite{MO}-\cite{BS}. 
The lattice QCD simulations show strong correlations 
between the instantons and the monopole 
currents~\cite{MO,MST,STSM}. 
Analytic studies of one and two instanton systems have been performed 
by making the abelian gauge fixing 
as the Polyakov gauge~\cite{SIST,STSM}. 
It has been shown that monopole trajectories tend to pass through 
instanton centers, while those far from instantons 
seem to be unstable under small variations.
Further in the recent studies, 
another type solution of the monopole trajectories are found to surround 
the instanton configurations in the maximally abelian gauge 

Theoretically, the relation between instantons and monopoles 
is suggested by the following consideration. 
The presence of a monopole can be signaled by a test Wilson loop 
surrounding it: its value is $-1$ if 
the monopole is inside the contour (we imply the $SU(2)$ 
color for simplicity). As for instantons, if one takes a 
Wilson loop whose one side is going through the center of 
an (anti)instanton and the other three sides are closing the loop at 
infinity, its value is also $-1$~\cite{DP}. This is true for any plane 
cutting through the center of an instanton. From this point of view, 
any instanton contains {\em spherically-distributed loops} of monopole 
currents winding around the 4-d center of the instanton. The length 
of the loop is typically given by the instanton size, $\rho$. 
When one makes an abelian gauge fixing of the instanton field 
to pinpoint the monopole loop inside the instanton, 
its concrete form and orientation in space depends on 
the particular abelian gauge fixing used and on the orientation 
of the instanton in color space. However, the presence of 
monopole loops inside instantons is, by itself, 
a gauge-invariant statement.

What happens with monopole trajectories when one takes an {\em
ensemble} of instantons ?  The answer to this question 
depends  on the ratio of the typical instanton size to the average
separation between instantons.  Let us introduce the dimensionless
{\it packing fraction} of instantons, $f=(\pi^2/2) \rho_0^4 N/V$, where 
$\rho_0$ is the most probable instanton size, $N$ is the total number of
pseudoparticles (instantons and anti-instantons) 
and $V$ is the 4-d volume. If this packing fraction is
small, the pseudoparticles are well separated. Hence, the monopole loops
associated with instantons do not mix. As the
packing fraction increases, the instantons start to contact each other,
and their monopole loops can well form large clusters going through
many instantons. This has been observed in an idealized
case~\cite{SIT} and in lattice investigations~\cite{HT,SSITA}. A
 quantitative question arises, at what packing fraction does the
``clusterization" happen, and what is the size of a typical cluster. If
clusters of the size of the physical volume appear, it may witness a
condensation of monopoles~\cite{Ezawa,Suz1}.

In principle, different intermediate situations may take place. 
For example, let the most probable size of instantons in the ensemble, 
$\rho_0$, be such that the packing fraction is, on the average, small. 
However,  if the size distribution has a long power-like tail 
at large sizes, certain rare but large instantons may overlap or just 
touch each other. As a result, the monopole loops inside them may hop 
from one pseudoparticle to another, so that a ``percolation''-type 
phenomenon takes place, with monopole trajectories forming 
large clusters. 

The present paper reports on the results of a numerical investigation 
on the relation between the clustering of monopoles and the instanton 
densities.

\vskip .7true cm

\noindent
{\bf 2. Instantons}

The field of one instanton with the size $\rho$ 
and the center $z_{\mu}$ in the singular gauge is 
\begin{equation}
A_{\mu}^{I}(x;z_k,\rho_k,O_k)
=\frac{i\rho^2{\tau^a}{O^{ai}}{\bar\eta}_{\mu\nu}^i(x-z)_\nu}
{(x-z)^2[(x-z)^2+\rho^2]},
\label{one}\end{equation}
where $O^{ai}$ is the instanton orientation matrix and 
$\bar\eta_{\mu \nu}^i$ is the 't~Hooft symbol~\cite{tH}. 
For {\em anti}-instantons 
$A_{\mu}^{\bar I}(x;z_k,\rho_k,O_k)$ one has to replace the 
$\bar\eta_{\mu \nu}^i$ symbol by 
$\eta_{\mu \nu}^i$. We shall work with the color $SU(2)$
group, hence the orientation matrix $O^{ai}$ is an orthogonal
3$\times$3 matrix which can be parametrized by 3 real parameters.
We use random orientation matrices for the ensemble of
instantons and anti-instantons (\IIs for short).

For simplicity we take a sum ansatz~\cite{Akas} 
for the system of \IIs, 
\begin{equation}
A_{\mu}(x)=\sum_k{A_{\mu}^I(x;z_k,\rho_k,O_k)}
+\sum_k{A_{\mu}^{\bar I}(x;z_k,\rho_k,O_k)}.
\label{sum}\end{equation}
An important point is the distribution in the sizes of instantons in
the ensemble. It has been recently shown analytically~\cite{DP} that
if the instanton size distribution falls off as $1/\rho^3$ at large
$\rho$, one gets a linear confining potential, with the string
tension proportional to the coefficient in the ``one over cube" law.
Previously certain arguments have been presented that the large-size
tail of the instanton size distribution should behave as
$1/\rho^5$~\cite{AS,Sh}. Therefore, in our investigation we take the
power of the tail $1/\rho^\nu$ of the instanton size distribution at
large $\rho$ to be $\nu=3$ or $\nu=5$. As for small sizes, the
distribution has to follow the one-loop result~\cite{tH}:  $d(\rho)\sim
\rho^{b-5}$ where $b=11N_c/3$ is the first Gell-Mann--Low
coefficient. 

We thus adopt the following size distribution,
\begin{equation}
d({\rho})={{1}\over{{({\rho\over{\rho_1}})^\nu}
+{({{\rho_2}\over\rho})^{b-5}}}}.
\label{sizedistr}\end{equation}
Here, $\rho_1$ and $\rho_2$ are certain parameters so
that the distribution \ur{sizedistr} is normalized to unity, while the
maximum of the distribution is fixed to a given value $\rho_0$, 
which is the most probable size of the
pseudoparticles in the ensemble.

\vskip .7true cm

\noindent
{\bf 3. Maximally Abelian Gauge Fixing}

We generate ensembles of $N$ pseudoparticles with
random orientations and centers; the size distribution is given by
Eq.(3). 
We take equal numbers of instantons and anti-instantons; 
$N_I=N_{\bar I}=N/2$. 
This is performed in the continuum theory. We then 
introduce a lattice and express the classical field in terms of the 
unitary matrices $U_\mu=\exp(iaA_\mu)$ living on the links. The 
continuum field is thus discretized. From now on, we do exactly the 
same as it is done in lattice calculations: we apply the maximally 
abelian gauge fixing~\cite{KLSW} to extract the monopole trajectories 
for each instanton ensemble. 
The only difference with the usual lattice calculations is that we 
start not from the ``hot" vacuum dominated by normal and 
probably innocent zero-point fluctuations, but from the ensemble 
of \IIs. It should be stressed, however, that our ensemble does not 
necessarily coincide with the ensemble of instantons obtained by 
cooling down the true or ``hot" Yang--Mills vacuum: 
the lattice cooling procedure does not only kill the 
zero-point oscillations but may also 
distort considerably the original ensemble of instantons, 
especially if \IIs happen to overlap.

We take a $16^4$ lattice with the lattice spacing of 
$a=0.15\rm fm$ and the most probable instanton size, 
$\rho_{0}=0.4\rm fm$. The volume is thus fixed and equal to 
$V=(2.4\rm fm)^4$. We consider four cases with the number of 
pseudoparticles 
equal to $N=$ 20, 40, 60 and 80, corresponding to the density 
$(N/V)^{1/4}=$ 174, 206, 228 and 245 MeV, respectively. 
Note that these values are to be compared with 200 MeV 
suggested by the study of instanton vacuum~\cite{Sh2,DP2}.
The corresponding packing fractions are 
$f=(\pi^2/2)\rho_0^4N/V=$ 0.076, 0.152, 0.228 and 0.305.
In all four cases we consider two regimes of the falloff of 
the size distributions: 
$1/\rho^3$ and $1/\rho^5$. The abelian gauge fixing is done by 
maximizing 
$R=\sum_{\mu,s} {\rm Tr}[U_\mu (s)\tau ^3U^{-1}_\mu (s)\tau ^3 ]$ 
in the maximally abelian gauge. 
We plot the distribution in the lengths of the monopole 
loops from 240 initial instanton ensembles. 

\vskip .7true cm

\noindent
{\bf 4. Numerical Results}

We show in Figs.1 and 2 the distributions in the lengths of monopole
loops. We see that for low
densities (Figs.1a and 2a) one has only relatively short monopole
loops, whose distribution follows the distribution in the sizes of
instantons \footnote{It is worth mentioning that the distribution of
relatively short monopole loops in true lattice QCD has been recently
measured by Hart and Teper who find a power behavior:
$P(L)\sim1/L^{2.87}$~\cite{HT2}. It indicates that in reality the
size distribution of instantons is rather $1/\rho^3$ than $1/\rho^5$.}.
It means that at low density there is no ``hopping" of monopole
trajectories from one instanton to another.  At larger densities the
length distribution starts to broaden.  Fig.~1b demonstrates the first
appearance of very large monopole clusters; in Fig.~1c their typical
length is already around 4000 lattice units, whereas the total number
of sites in our 16$^4$ lattice is 65,536.  It means that about $1/16$
of the total number of sites is connected by monopole currents.  Since
$1/16 \ll 1$ we are not having problems in identifying monopole
currents. At the same time a 4000-link cluster actually covers the
whole physical volume. We conclude that somewhere between Figs.~1b 
and 1c a percolation-type phase transition occurs. Note that
it occurs when the bulk of instantons are still relatively dilute
(the average packing fraction is between 1/6 and 1/4), and the
instanton density is of the order of $(200\rm MeV)^4$ 
which is a very reasonable value from the point of view 
of the instanton vacuum~\cite{Sh2,DP2}.
The same phenomenon happens in the $1/\rho^5$ case, see Fig.~2d, but
at instanton densities about twice larger than in the $1/\rho^3$ case,
which is probably less realistic.

It is instructive to compare our results with those of 
the true $SU(2)$ lattice QCD with $16^3 \times 4$ 
at different temperatures~\cite{SSITA,Suz1} as shown Fig. 3. 
We see that the distribution of the monopole lengths 
in the confinement phase ($\beta=2.2$) resembles the 
distribution we get from the instanton ensemble 
at the density around $(228\rm~MeV)^4$.
The distribution obtained in the deconfinement phase is 
very similar to that which we get from a low-density instanton ensemble.
Since the density of instantons is known to be suppressed at high 
temperatures, the deconfinement phase transition can be thus 
qualitatively understood as due to the 
inevitable dilution of instantons at high temperatures.

To conclude, we have demonstrated that a random ensemble of
instantons and anti-instantons produces, after the standard maximally
abelian gauge fixing, enormous clusters of monopole loops. This is 
quite similar to what is obtained in the true lattice QCD, where the 
appearance of enormous clusters is 
believed to have relevance to the confining behavior of the Wilson
loop. The percolation-like phase transition to monopole clusters
covering the whole physical volume happens when the bulk of
instantons is still relatively dilute (their packing fraction is
from 1/6 to 1/4). It should be stressed that our ensemble of
instantons has much entropy due to random positions, sizes and
orientations; cooling such an ensemble would probably freeze out
most of its degrees of freedom. That might explain why the well-cooled
lattice where instantons are usually detected, looses most of its
original string tension.

\vskip .7true cm

\noindent
{\bf Acknowledgment}

D.~D. has been supported in part by a Fellowship in the Priority
Research Area from the Japan Society for the Promotion of Sciences
and acknowledges warm hospitality at the RCNP in Osaka.
We are grateful to V.~Petrov and F.~Araki for useful discussions.

\vskip .7true cm

\begin{center}
\large\bf Figure Captions
\end{center}

\item{\rm Fig.1}:  Histograms of monopole loop lengths for the number of 
pseudoparticles ; $N=$20, 40, 60 and 80, which correspond to the 
instanton densities of $(N/V)^{1/4}=$174, 206, 228 and 245 MeV, 
respectively. This is the case of the instanton size distribution 
with the fall off as $1/\rho^3$, $\nu=3$ in Eq.(3).

\item{\rm Fig.2}:  Histograms of monopole loop lengths for the number of 
pseudoparticles ; $N=$20, 40, 60 and 80. The numbers indicated are the 
corresponding instanton 
densities ; $(N/V)^{1/4}$.
The instanton size distribution falls off as 
$1/{\rho^5}$, $\nu=5$ in Eq.(3).

\item{\rm Fig.3}:  The comparison of the distributions of 
the monopole loop length. 
Shown in the upper part are the $SU(2)$ lattice QCD results at 
$\beta=2.35$ and $\beta=2.2$, which correspond to the deconfinement 
(high temperature) and the confinement (low temperature) phases, 
respectively~\cite{SSITA}. The lower part denotes the results for the 
instanton ensemble with the instanton size distribution, 
$\nu=3$, at the instanton densities 
$(N/V)^{1/4}=$174 MeV (low instanton density) and 
$(N/V)^{1/4}=$228 MeV (high instanton density).


\begin{thebibliography}{}

\bibitem{Nambu}
Y.~Nambu, Phys.~Rev.~{\bf D10} (1974) 4262. \\
G.~'t~Hooft, in {\it High Energy Physics} 
(Editorice Compositori, Bologna 1975). \\
S.~Mandelstam, Phys.~Rep.~{\bf C23} (1976) 245.

\bibitem{tHmono}
G.'t~Hooft, Nucl. Phys. {\bf B190} (1981) 455. 

\bibitem{Ezawa}
Z.~F.~Ezawa and A.~Iwazaki, Phys.~Rev.~{\bf D25} (1982) 2681; 
{\bf D26} (1982) 631.

\bibitem{tH}
G.'t~Hooft, Phys. Rev. Lett. {\bf 37} (1976) 8; 
Phys. Rev. {\bf D14} (1976) 3432.

\bibitem{Akas}
D.~Diakonov, 
in {\it Gauge Theories of the Eighties}, 
Lecture Notes in Physics, (Springer-Verlag, 1983) 207.

\bibitem{D1}
D.~Diakonov, 
Lectures at the Enrico Fermi School (1995), 
in press; hep-ph/9602375.

\bibitem{PV}
M.~Polikarpov and A.~Veselov,  Nucl. Phys. {\bf B297} (1988) 34.

\bibitem{CDMPV}
M.~Campostrini, A.~Di~Giacomo, M.~Maggiore, H.~Panagopou\-los 
and E.~Vi\-cari,  Nucl. Phys. {\bf B329} (1990) 683.

\bibitem{CGHN}
M.-C.~Chu, J.~Grandy, S.~Huang and J.~Negele, Phys. Rev. Lett.
{\bf 70} (1993) 225;  Phys. Rev. {\bf D49} (1994) 6039.

\bibitem{MO}
O.~Miyamura and S.~Origuchi, Proc. of Int. Workshop on 
{\it Color Confinement and Hadrons},
Osaka, edited by H.~Toki et al. (World Scientific,1995) 235.

\bibitem{SIST}
H.~Suganuma, H.~Ichie, S.~Sasaki and H.~Toki, 
in {\it Color Confinement and Hadrons} [7] 
(World Scientific,1995) 65. \\
H.~Suganuma, K.~Itakura, H.~Toki and O.~Miyamura, 
Proc. of Int. Workshop on {\it Nonperturbative Approaches to QCD}, 
Trento, edited by D.~Diakonov, (PNPI, 1995) 224.

\bibitem{MST}
S.~Thurner, H.~Markum and W.~Sakuler, 
in {\it Color Confinement and Hadrons} [7] (World Scientific, 1995) 77. \\
H.~Markum and W.~Sakuler and S.~Thurner, 
Nucl. Phys. {\bf B} (Proc. Suppl.) {\bf 47} (1996) 254.

\bibitem{CG}
M.N.~Chernodub and F.V.~Gubarev, JETP Letter {\bf 62} (1995) 100;
in {\it Nonperturbative Approaches to QCD} [8] (PNPI, 1995) 217.

\bibitem{STSM} 
H.~Suganuma, A.~Tanaka, S.~Sasaki and O.~Miyamura, 
Nucl. Phys. {\bf B} (Proc. Suppl.) {\bf 47} (1996) 302.

\bibitem{HT}
A.~Hart and M.~Teper, Phys. Lett. {\bf B371} (1996) 261.

\bibitem{BS}
V.~Bornyakov and G.~Schierholz, preprint; hep-lat/9605019.

\bibitem{DP}
D.~Diakonov and V.~Petrov, 
in {\it Nonperturbative Approaches to QCD} [8] (PNPI, 1995) 239.

\bibitem{SIT}
H.~Suganuma, K.~Itakura and H.~Toki, hep-th/9512141.

\bibitem{SSITA}
H.~Suganuma, S.~Sasaki, H.~Ichie, H.~Toki and F.~Araki, 
Proc. of Int. Symp. on {\it Frontier '96}, Osaka,
Mar. 1996, (World Scientific) in press. 

\bibitem{Suz1}
S.~Kitahara, Y.~Matsubara and T.~Suzuki, Prog. Theor. Phys.
{\bf 93} (1995) 1. \\ 
T.Suzuki, 
in {\it Nonperturbative Approaches to QCD} [8] (PNPI, 1995) 51.

\bibitem{AS}
N.~Agasyan and Yu.~A.~Simonov, Mod. Phys. Lett. {\bf A10} (1995)
1755.

\bibitem{Sh}
E.~Shuryak, Phys. Rev. {\bf D52} (1995) 5370.

\bibitem{KLSW}
A.S.~Kronfeld, M.L.~Laursen, G.~Schierholz and U.-J.~Wiese,
Phys. Lett. {\bf 198} (1987) 576.

\bibitem{HT2}
A.~Hart and M.~Teper, Proc. of {\it Lattice '96}, St. Louis, 
Nucl. Phys. {\bf B} in press; hep-lat/9606022.

\bibitem{Sh2}
E.~Shuryak, Nucl. Phys. {\bf B203} (1982) 93.

\bibitem{DP2}
D.~Diakonov and V.~Petrov, Nucl. Phys. {\bf B245} (1984) 259. \\
D.~Diakonov, M.~Polyakov and C.~Weiss,  Nucl. Phys. {\bf B461}
(1995) 539.



\end{thebibliography}
\end{document}